\documentclass[12pt, letterpaper]{JHEP3}

\usepackage{amssymb, amsmath, amsopn, amsthm}
\usepackage{latexsym}
\usepackage{graphics}
\usepackage{epsfig}

\title{Taming the Tachyon\\ in Cubic String Field Theory}

\author{Erasmo Coletti, Ilya Sigalov, Washington Taylor \\
{Center for Theoretical Physics} \\ {Massachusetts Institute of Technology} \\
{Cambridge, MA 02139, U.S.A.} \\ {\tt colettie@mit.edu,
sigalov@mit.edu, wati\ {\rm at}\ mit.edu}}

\abstract{We give evidence based on level-truncation computations that
the rolling tachyon in cubic open string field theory
(CSFT) has a
well-defined but wildly oscillatory time-dependent solution which goes
as $e^t$ for $t \rightarrow -\infty$.  We show
that a field redefinition taking the CSFT effective tachyon action to
the analogous boundary string field theory (BSFT) action takes the
oscillatory CSFT solution to the pure exponential solution $e^t$ of the
BSFT action.}

\preprint{hep-th/0505031, MIT-CTP-3658}

\newcommand{\G}{\Gamma}
\newcommand{\pd}{\partial}

\newcommand{\half}{\frac{1}{2}}

\newcommand{\cP}{\mathcal{P}}
\newcommand{\cQ}{\mathcal{Q}}

\newcommand{\cU}{\mathcal{U}}
\newcommand{\cV}{\mathcal{V}}

\newcommand{\cX}{\mathcal{X}}

\newcommand{\td}{\tilde}

\DeclareMathOperator{\Det}{Det}

\begin{document}
\section{Introduction}

The tachyon of the open bosonic string has played an important role
in recent years in the development of string field theory as a
background-independent formulation of string theory.  Following Sen's
conjectures regarding this tachyon \cite{Sen-conjectures}, significant
progress has been made towards demonstrating that both the unstable vacuum
containing the tachyon and the ``true'' vacuum where the tachyon has
condensed are well-defined states in Witten's cubic open
string field theory (CSFT) \cite{Witten-CSFT}.  This is important evidence
that string field theory is capable of describing multiple distinct
vacuum configurations using a single set of degrees of freedom, as one
would expect for a background-independent formulation of the theory.
Some of the work in this area is reviewed in \cite{Taylor-Zwiebach,Sen-review}.

An important aspect of the open string tachyon which is not yet
fully understood, however, is the dynamical process through which
the tachyon rolls from the unstable vacuum to the stable vacuum.  A
review of previous work on this problem is given in
\cite{Sen-review}. Computations using CFT, boundary states, RG flow
analysis and boundary string field theory (BSFT)
\cite{Witten:1992qy, Witten:1992cr, Shatashvili:1993kk,
Shatashvili:1993ps} show that the tachyon should monotonically roll
towards the true vacuum, but should not arrive at the true vacuum in
finite time \cite{Sen:2002nu}-\cite{lnt}. In BSFT variables, where
the tachyon $T$ goes to $T \rightarrow \infty$ in the stable vacuum,
the time-dependence of the tachyon field goes as $T (t) =e^t$.  This
dynamics is intuitively fairly transparent, and follows from the
fact that $e^{t}$ is a marginal boundary operator
\cite{et1,et2,et3,Sen:2002nu,lnt}. Other approaches to understanding
the rolling tachyon from a variety of viewpoints including DBI-type
actions \cite{DBI1}-\cite{DBIn}, S-branes and timelike Liouville
theory \cite{Liouville1}-\cite{Liouvillen}, matrix models
\cite{mm1}-\cite{mmn}, and fermionic boundary CFT \cite{fermion}
lead to a similar picture of the time dynamics of the tachyon.

In CSFT, on the other hand, the rolling tachyon dynamics appears
much more complicated.  In \cite{Moeller-Zwiebach}, Moeller and
Zwiebach used level truncation to analyze the time dependence of the
tachyon. They found that at low levels of truncation, the tachyon
rolls well past the minimum of the potential, then turns around and
begins to oscillate with ever increasing amplitude. It was further
argued by Fujita and Hata in \cite{Fujita-Hata} that such
oscillations are a natural consequence of the form of the CSFT
equations of motion, which include an exponential of time
derivatives acting on the tachyon field.

These two apparently completely different pictures of the tachyon
dynamics raise an obvious puzzle.  Which picture is correct?  Does the
tachyon converge monotonically to the true vacuum, or does it undergo
wild oscillations?  Is there a problem with the BSFT approach?  Does
the CSFT analysis break down for some reason such as a branch point
singularity at a finite value of  $t$?  Does the dynamics in
CSFT behave better when higher-level states are included?  Is
CSFT an incomplete formulation of the theory?

In this paper we resolve this puzzle.  We carry out a systematic
level-truncation analysis of the tachyon dynamics for a particular
solution in CSFT.  We compute the trajectory $\phi (t)$ as a power
series in $e^t$ at various levels of truncation.  We show that
indeed the dynamics in CSFT has wild oscillations.  We find,
however, that the trajectory $\phi (t)$ is well-defined in the sense
that increasing both the level of truncation in CSFT and the number of
terms retained in the power series in $e^t$ leads to a convergent
value of $\phi (t)$
for any fixed $t$, at least below an upper bound $t < t_b$
associated with the limit of our computational ability.

We reconcile this apparent discrepancy with the results of BSFT by
demonstrating that a field redefinition which takes the CSFT action
to the BSFT action also maps the wildly oscillating CSFT solution to
the well-behaved BSFT exponential solution.  This qualitative change
in behavior through the field redefinition is possible because the
field redefinition relating the tachyon in the two formulations is
nonlocal and includes terms with arbitrarily many time derivatives.
Such field redefinitions are generically expected to be necessary
when relating the background-independent CSFT degrees of freedom to
variables appropriate for a particular background
\cite{Ghoshal-Sen}. A similar field redefinition involving higher
derivatives was shown in \cite{cst} to be necessary to relate the
massless vector field $\hat{A}_\mu$ of CSFT on a D-brane with the
usual gauge field $A_\mu$ appearing in the Yang-Mills and
Born-Infeld actions. Other approaches to the rolling tachyon using
CSFT appear in \cite{otherCSFT1}-\cite{otherCSFTn}; related
approaches which have been studied include $p$-adic SFT
\cite{Yang:2002nm,Moeller:2003gg}, open-closed SFT \cite{oc}, and
vacuum string field theory \cite{VSFT,VSFTn}. Closed string
production during the rolling process is described in
\cite{Okuda:2002yd, Lambert:2003zr, Shelton:2004ij}.

The paper is organized as follows.  Section \ref{sec:solvingeoms}
describes the general approach that we use to find the rolling tachyon
solution and gives the leading order terms in the solution
explicitly.  Section \ref{sec:numerical} describes the results of
numerically solving the equations of motion in level-truncated CSFT.
Section \ref{sec:redef} is dedicated to finding the leading terms in
the field redefinition that relates the effective tachyon actions in
Boundary and Cubic String Field Theory.  Section \ref{sec:discussion}
contains conclusions and a discussion of our results.
Some technical details regarding our methods of calculation are
relegated to Appendices.

As this paper was being completed the paper \cite{Forini:2005bs}
 appeared, which treats the same system, although without considering
 massive fields.  The analysis of \cite{Forini:2005bs} is carried out
 using analytic methods which give an approximate rolling tachyon
 solution when all fields other than the tachyon are neglected.  The
 solution in their paper shares  some qualitative features with our
 results---in particular, they find a solution which has similar
 behavior for negative time, and their solution also rolls past the
 naive minimum of the tachyon potential.  Their solution has a cusp at
 $t = 0$ where the solution has a discontinuous first derivative; we
 believe that their solution breaks down at this point, but that
 their solution is good for $t < 0$ and that the analytic methods they
 use in deriving their results are of interest and may help
in understanding the dynamics of the system.

\section{Solving the CSFT equations of motion}

\label{sec:solvingeoms}

 We are interested in finding a solution to
the complete open string field theory equations of motion.  The full
CSFT action contains an infinite number of fields, coupled through
cubic terms which contain exponentials of derivatives (see
\cite{Taylor-Zwiebach} for a detailed review).  Thus, we have a
nonlocal action in which it is difficult to make sense of an initial
value problem (see
\cite{Brekke:dg,Volovich:2003zh,Vladimirov:2003kg, Gomis:2003xv} for
some discussion of such equations with infinite time derivatives).

Nonetheless, we can systematically develop a solution valid for all
times by assuming that as $t \rightarrow -\infty$ the solution
approaches the perturbative vacuum at $\phi = 0$.  In this limit the
equation of motion is the free equation for the tachyon field
$\ddot{\phi} (t)= \phi (t)$, with solution $\phi (t) = c e^{t}$. For
$t\ll 0$, we can perform a perturbative expansion in the small
parameter $e^{t}$.
The computations carried out in this paper indicate that this power
series indeed seems convergent for all $t$. A related approach was
taken in \cite{Moeller-Zwiebach, Fujita-Hata}, where an expansion in
$\cosh t$ was proposed. This allows a one-parameter family of
solutions with $\dot{\phi} (0) = 0$, but is more technically
involved due to the more complicated structure of $\cosh nt$
compared with $e^{nt}$. We restrict attention here to the simplest
case of solutions which can be expanded in $e^{t}$, but we expect
that a more general class of solutions can be constructed using this
approach.  Note that in most previous work on this problem,
solutions have been constructed using Wick rotation of periodic
solutions; in this paper we work directly with the real solution
which is a sum of exponentials.

The infinite number of fields of CSFT represents an additional
complication.  We can, however, systematically integrate out any
finite set of fields to arrive at an effective action for the tachyon
field which we can then solve using the method just described.  We do
this using the level-truncation approximation to CSFT including
fields up to a fixed level.  We find that the resulting trajectory
$\phi (t)$ converges well for fixed $t$ as the level of truncation is
increased.

We thus compute the solution $\phi (t)$ with the desired behavior
$e^{t}$ as $t \rightarrow -\infty$ in two steps.  In the first step,
described in subsection \ref{sec:effaction}, we compute the tachyon
effective action, eliminating all the other modes using equations of
motion.  Some technical details of this calculation are relegated to
Appendix A.  In the second step, described in subsection
\ref{sec:solvingeom}, we write down the equation of motion for the
effective theory and solve it perturbatively in powers of $e^{t}$.

\subsection{Computing the effective action}

\label{sec:effaction}
We are interested in a spatially homogeneous rolling tachyon
solution. One can compute such a solution by solving the equations
of motion for the infinite family of string fields with all the
spatial derivatives set to $0$. Labeling string fields $\psi_i$, the
cubic string field theory equations of motion (in the Feynman-Siegel
gauge) take the schematic form
\begin{equation}
(\partial_t^2 -m_i^2) \psi_i (t) = g\, e^{ V^{11}_{00} (\pd^2_{s} +
\pd^2_u + \pd_s \pd_u)} C_{i}^{jk}(\pd_u, \pd_s)\ \psi_j (s) \psi_k
(u)|_{s = u = t} \label{eq:all-equations}
\end{equation}
where all possible pairs of fields appear on the RHS. The
coefficients $C_{ijk}$ multiplying each term may contain a finite
order polynomial in the derivatives $\partial_s, \partial_u$.
Plugging in the Ansatz $\phi (t) = \psi_0 (t) = e^t + \cdots$ with
all other fields vanishing at order $e^t$ it is clear that we can
systematically solve the equations for all fields order by order in
$e^t$.  This is one way of systematically solving order by order for
$\phi (t)$.

We will find it convenient to think of the perturbative solution for
$\phi (t)$ in terms of an effective action $S [\phi]$ which arises
by integrating out all the massive string fields at tree level.
Perturbatively, we can solve the equations of motion
(\ref{eq:all-equations}) for all fields except $\phi = \psi_0$ as
power series in $\phi$, by recursively plugging in the equations of
motion for all fields except $\phi$ on the RHS until all that
remains is a perturbative expansion in terms of $\phi (t)$ and its
derivatives.  We have used two approaches to compute the effective
action $S [\phi]$.  One approach is to explicitly use the equations
(\ref{eq:all-equations}) for all fields up to a fixed level.  This
approach is useful for generating terms to high powers in $g$ but
becomes unwieldy for fields at high levels.  The second approach we
use is to compute the effective action as a diagrammatic sum using
the level truncation on oscillator method developed in
\cite{WT-perturbative}.  This approach is useful for calculating
low-order terms in the effective potential where high-level fields
are included.  Some details of the oscillator approach are described
in Appendix A.

The leading terms in the tachyon action are the quadratic and cubic
terms coming directly from the CSFT action
\begin{equation}
S [\phi] = \frac{1}{2}\int d t \; \phi (t) \left(-\partial_t^2 +1
\right) \phi (t) -\frac{g}{3}\left(e^{\frac{1}{2} V^{11}_{00} (
\partial_t^2 -1) } \phi (t) \right)^3 + \cdots \label{eq:action-cubic}
\end{equation}
where
\begin{equation}
V^{11}_{00} = - \log\left(\frac{27}{16}\right)
\end{equation}
is the Neumann coefficient for the three tachyon vertex.

Integrating out the massive fields at tree level gives rise to
higher-order terms $g^2 \phi^4, \ldots$ with even more complicated
derivative structures.    The resulting effective action can be
written in terms of the (temporal) Fourier modes $\phi (w)$ of $\phi
(t)$ as
\begin{equation}
\label{eq:effaction} S [\phi] = \sum_n \frac{g^{n-2}}{n!} \int
\prod_{i=1}^n d w_i \; (2 \pi)^n \delta\Bigl(\sum_i w_i\Bigr)
\Xi_n^{\text{CSFT}}(w_1, \dots, w_n) \phi (w_1)  \dots \phi(w_n)
\end{equation}
where the functions $\Xi_n^{CSFT}(w_1, \dots, w_n)$ determine the
derivative structure of the terms at  order $ g^{n-2}\phi^n$.  The
quadratic and cubic terms following from (\ref{eq:action-cubic}) are
\begin{align}
\label{eq:quadcsftXi}
\Xi^{\text{CSFT}}_2 (w_1, w_2) &= (1 - w_1 w_2), \\
\label{eq:cubcsftXi} \Xi^{\text{CSFT}}_3 (w_1, w_2, w_3) &= - 2\,
e^{-\half V^{11}_{00} (w_1^2 + w_2^2 + w_3^2 +3)}.
\end{align}

One way to obtain the approximate classical effective action for the
tachyon field is to use the equations of motion for a few low level
massive fields to eliminate these fields explicitly from the action.
The higher level massive fields are set to zero (level truncation).

As an example, we now explicitly compute the quartic term in the
effective action (\ref{eq:effaction}) in the level 2 truncation. In
the case of CSFT for a single D-brane the combined level of fields
coupled by a cubic interaction must be even. For example, there is
no vertex coupling two tachyons (level zero) with the gauge boson
(level 1). It follows that there are no tree level Feynman diagrams
with all external tachyons and internal fields of odd level. Thus,
in calculating the tachyonic effective action we may set odd level
fields to 0. Fixing Feynman-Siegel gauge, the only fields involved
are the tachyon $\phi$ and three level 2 massive fields with $m^2 =
1$: $\beta$, $B_\mu $ and $B_{\mu\nu}$. The terms in the action
contributing to the four-tachyon term in the effective action are
\begin{multline}
\label{eq:L2cubic} \half \int dt\, \beta (\pd_t^{2}+1)\beta -
B_{\mu\nu}(\pd_t^{2}+1)B^{\mu\nu} - B_{\mu}(\pd_t^{2}+1)B^{\mu}
+\\
g \int dt \, a_1 \td \phi^{2}\td B_{\mu}^{\mu}+ a_2
(\td\phi\pd_{t}\pd_{t}\td\phi- \pd_{t}\td\phi\pd_{t}\td\phi ) \td
B^{00}+ a_3\td\phi^{2}\td\beta + a_4\td\phi\pd_{t} \td\phi \td B^{0}
\end{multline}
where $\td f=e^{\half V_{00}^{11}(\pd^{2}_t-1)}f$. Other interaction
terms involving level 2 fields, for example $\beta^3$ or $B_\mu
B^\mu \phi$, would contribute to the effective action at higher
powers of $\phi$. The coefficients $a_{1}$, ... $a_4$ are real
numbers and can be expressed via the appropriate matter and ghost
Neumann coefficients,
\begin{align}
a_1 &= -\frac{V_{11}^{11}}{\sqrt{2}}\approx 0.130946, & a_2 &= \sqrt{2} (V_{01}^{12})^{2} \approx 0.419026, \nonumber\\
a_3 & = X_{11}^{11}\approx  0.407407, & a_4 & = - 6
V_{02}^{12}\approx 0.628539.
\end{align}

Following the procedure described above we write down the equations
of motion for the massive fields, and plug them into
(\ref{eq:L2cubic}). We then obtain a quartic term in the tachyonic
effective action,
\begin{multline}\label{eq:L2quartic}
g^2 e^{-3V_{00}^{11}} \int \prod_{i = 1}^{4} {(2 \pi d w_i)}
\phi(w_{i})\delta(\sum w_i) \frac{\exp\big(- V_{00}^{11}
[w_{1}^{2}+w_{2}^{2}+w_{3}^{2}+w_{4}^{2}+w_{1}w_{2}+w_{3}w_{4}]\big)}
{1 - (w_{1}+w_{2})^2 }\\
\Bigl( b_{1} + b_{2}\,w_{2}(w_2 - w_1) + b_{3}\,w_1 w_4 (w_2-w_1)
(w_4 -w_3) + b_4\, w_2 w_4 \Bigr),
\end{multline}
where we have denoted
\begin{align}
b_{1}&=\half(13 (V_{11}^{11})^{2}- (X_{11}^{11})^{2}),
&b_{2}&= -V_{11}^{11}(V_{01}^{12})^{2}, \nonumber\\
b_{3}&=(V_{01}^{12})^{4}, & b_{4}&= 18 (V_{02}^{12})^{2}.
\end{align}
We have explicitly computed the terms to order $\phi^7$ in the
effective tachyon action in the $L=2$ truncation.  One can in
principle continue the procedure further, increasing both the level of
truncation and the powers of $\phi$ in the effective action.  Explicit
calculation, however, becomes laborious as we take into account more
and more string field components; the oscillator method
\cite{WT-perturbative}, described in the appendix \ref{app:A} is more
efficient for high-level computations. In the next section we
proceed to find the solution of the equations of motion
from the effective tachyon action.

\subsection{Solving the equations of motion in the effective theory}
\label{sec:solvingeom}

We now outline the process for solving the tachyon equation of
motion for the effective theory, and we compute the first
perturbative corrections to the free solution.
The variation of the effective action (\ref{eq:effaction}) gives
equations of the form
\begin{equation}
\label{eq:eomgeneric} (\pd_t^2-1) \phi = \sum_{n = 2}^\infty g^{n-1}
K_{n}(\phi, \dots, \phi)
\end{equation}
where the nonlinear terms of order $\phi^{n}$ are denoted by $K_n$.
The specific form of the $K_n$ follow by differentiating
(\ref{eq:effaction}) with respect to $\phi (t)$.  The functions
$\Xi_n$ appearing in (\ref{eq:effaction}), and thus the
corresponding $K_{n -1}$'s can in principle be explicitly computed
for arbitrary $n$ at any finite level of truncation using the method
described in the previous subsection.  An alternate approach which
is more efficient for computing $K_n$ at small $n$ but large
truncation level is reviewed in  Appendix A.  In general,
independently of the method used to compute it, $K_n$ will be a
complicated momentum-dependent function of its arguments.

The solution of the linearized equations of motion which satisfies
the boundary condition $\phi \rightarrow 0$ as $t \rightarrow
-\infty$ is $\phi(t) = c_1 e^t$. As discussed above, we wish to use
perturbation theory to find a rolling solution which  is defined by
this asymptotic condition as $ t \rightarrow -\infty$.  Note that
this asymptotic form places a condition on all derivatives of $\phi$
in the limit $t \rightarrow -\infty$, as appropriate for a solution
of an equation with an unbounded number of time derivatives.  If we
now assume that the full solution can be computed by solving
(\ref{eq:eomgeneric}) using perturbation theory, at least in some
region $t<t_{\text{max}}$, it can be easily seen that the successive
corrections to the asymptotic solution $\phi_1 (t) = c_1 e^{t}$ are
of the form $\phi_n(t) = c_{n} e^{nt}$. In other words, to solve the
equations of motion using perturbation theory we expand $\phi(g, t)$
in powers of $g$
\begin{equation}
\label{eq:phiexpansion}
\phi(g, t) = \phi_1(t) + g \phi_2(t) + g^2 \phi_3(t)
+ \dots
\end{equation}
where
\begin{equation}
\phi_{n}(t) = c_n e^{nt}.
\label{eq:phi-c}
\end{equation}
As we will see, our assumption leads to a power series which seems
to be convergent for all $t$ and all $g$.  Note that since $g^n e^{n
t} = e^{n(t+\log(g)) }$, the coupling constant can be set to 1 by
translating the time variable and rescaling $\phi$, so convergence
for fixed $g$ and all $t$ implies convergence for all $t$ and for
all $g$. Plugging (\ref{eq:phiexpansion}) into (\ref{eq:eomgeneric})
we find
\begin{equation}
\label{eq:EOM-expanded} (\pd_t^2-1) \phi_n =  (n^2-1) c_n e^{nt} =
\sum_p \sum_{m_1 + m_2 + \dots m_p = n} K_{p}(\phi_{m_1},
\phi_{m_2}, \dots , \phi_{m_p}).
\end{equation}
These equations allow us to solve for $c_{n > 1}$ iteratively in $n$.
Having solved the equations for $c_2, \ldots , c_{n-1}$ we can plug
them in via (\ref{eq:phi-c}) on the right hand side of
(\ref{eq:EOM-expanded}) to determine $c_n$.

As an example, let us consider the first correction $ \phi_2 (t) =c_2
e^{2t}$ to the linearized solution $\phi_1(t) = c_1 e^{t}$.  The
equation of motion at quadratic order arising from $K_2$ is
\begin{equation}
\label{eq:eom1} (\pd_t^2-1) \phi = - e^{\half V^{11}_{00} ( \pd_t^2
- 1)} (e^{\half V^{11}_{00} (\pd_t^2 - 1)} \phi)^2.
\end{equation}
Plugging in $\phi_1 = c_1 e^t$, $\phi_2 = c_2 e^{2 t}$ we find
\begin{equation}
c_2 (\pd_t^2-1) e^{2 t} = -c_1^2 e^{\half V^{11}_{00} (\pd_t^2-1)}
(e^{\half V^{11}_{00} (\pd_t^2 - 1)} e^t)^2
\end{equation}
and therefore
\begin{equation}
c_2 = -\frac{1}{3} e^{\frac{3}{2} V^{11}_{00}} c_1^2.
\label{eq:answer-c2}
\end{equation}
If we normalize $c_1 = 1$ then the solution to order $e^{2t}$ is
\begin{equation}
\phi(t) = e^t - \frac{64}{243\,{\sqrt{3}}}e^{2\, t} +\dots\ .
\end{equation}
The quartic interaction term in the effective action would
contribute to coefficients $c_n e^{nt}$ with $n\ge 3$ with the
leading order contribution being $c_3 e^{3t}$. From equation
(\ref{eq:EOM-expanded}) we have
\begin{equation}
\label{eq:c3eq} c_3  = \frac{e^{-3 t}}{8} \left( 2 c_2 K_2(e^t,
e^{2t}) + K_{3}(e^t, e^t, e^t) \right).
\end{equation}
where $K_3$ is obtained by differentiating (\ref{eq:L2quartic}) with
respect to $\phi(t)$. The two summands in (\ref{eq:c3eq}) represent
contributions from the cubic and quartic terms in the effective
action. The numerical values of these contributions are
\begin{align}
(\delta c_3)_{\text{cubic}}  &\cong 0.0021385, &(\delta
c_3)_{\text{quartic}}
 &\cong 0.0000492826.
\end{align}
It is perhaps surprising that the contribution to $c_3$ from the
quartic term in the effective action is merely $0.2\%$ of the
contribution from the cubic term. Adding the contributions we get
the rolling solution to second order in perturbation theory in level
2 truncation
\begin{equation}
\phi(t) \cong e^t - 0.152059 e^{2\,t} + 0.002187\,e^{3\,t} + \dots\
.
\end{equation}
In this section we have explicitly demonstrated our procedure for
the calculation of the rolling tachyon solution. We considered the
CSFT action truncated to fields with level less or equal than two
and computed the first two corrections to the solution of the
linearized equations of motion. The next section is dedicated to the
more detailed numerical analysis of the rolling tachyon solution.

\section{Numerical results}
\label{sec:numerical}

In this section we describe the results of using the level-truncated
effective action $S[\phi]$ to compute approximate perturbative
solutions to the equation of motion through (\ref{eq:EOM-expanded}).
We are testing the convergence of the solution in two respects.  In
subsection \ref{sec:couplingup} we check that the solution converges
nicely at fixed $t$ when we take into account successively higher
powers of $\phi$ in a perturbative expansion of the effective action
while keeping the truncation level fixed at $L = 2$.  In subsection
\ref{sec:levelup} we check that the solution converges well for
fixed $t$ when we keep the order of perturbation theory fixed while
increasing the truncation level.

\subsection{Convergence of perturbation theory at $L = 2$}
\label{sec:couplingup}

The equation (\ref{eq:EOM-expanded}) allows us to find the successive
perturbative contributions to the solution of the equations of motion,
given an explicit expression for the terms in the effective action.
The solution takes the form
\begin{equation}
\phi(t) = \sum_n c_n e^{n t}.
\label{eq:expand-solution}
\end{equation}
Since all the derivatives of $e^{nt}$ are straightforward to
compute, as in (\ref{eq:answer-c2}), we can replace these
derivatives in any operator through $f (\partial_t) e^{nt}
\rightarrow f (n) e^{nt}$. This manipulation is justified as long as
$f$ is regular at $n$.

We have computed the functions $\Xi_n^{{\rm CSFT}}$ and the
resulting $K_{n -1}$'s by solving the equations of motion up to $n =
7$ and integrating out all fields at truncation level $L = 2$ as
described in subsection 2.1. We have used these $K_n$'s to compute
the resulting approximate coefficients $c_n$, with $n \le 6$.  To
compute the coefficient $c_n$ one needs the effective tachyon action
computed to order $n+1$; higher terms in the action contribute only
to higher order coefficients.  \FIGURE {
\includegraphics{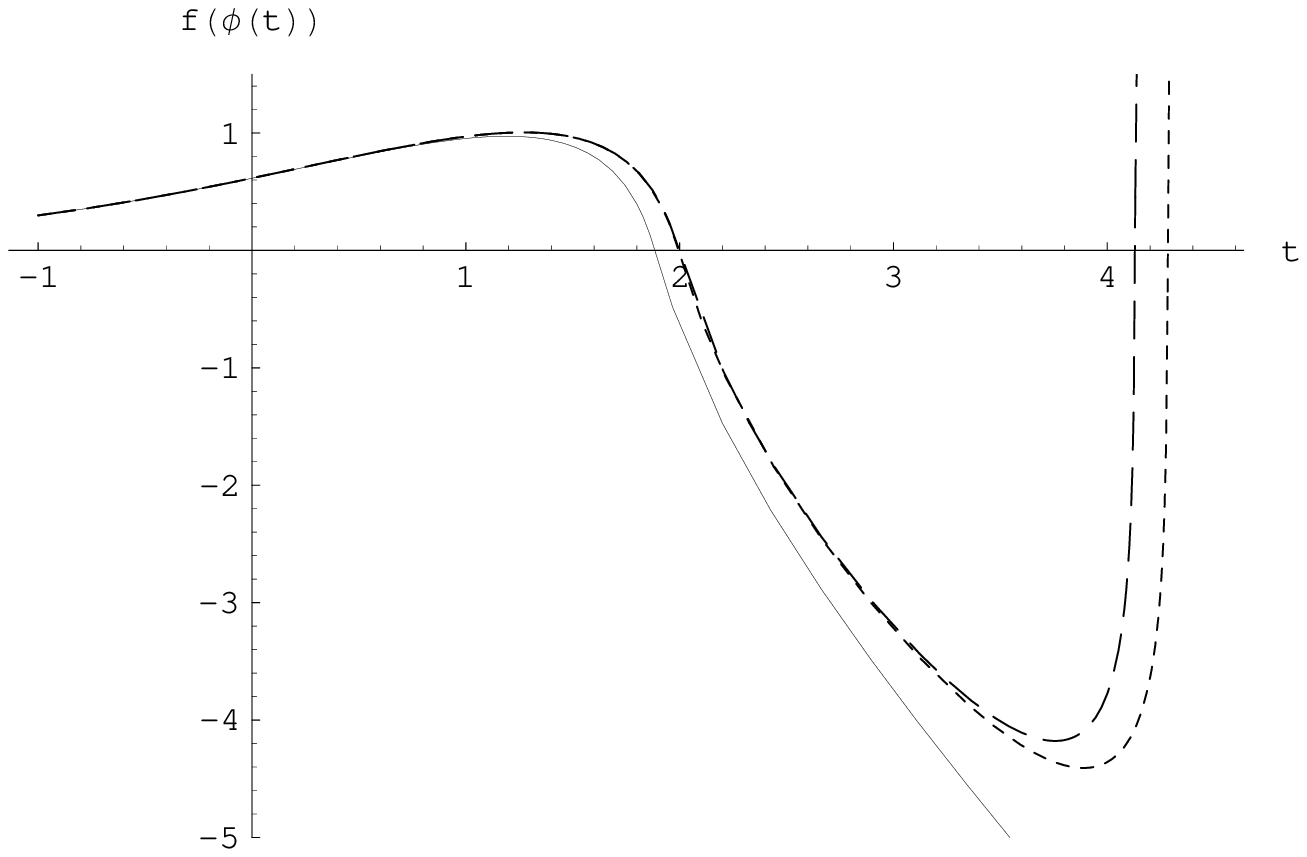}
\caption[x]{\footnotesize The solution $\phi(t)$ including the first
two turnaround points, including fields up to level $L = 2$. The
solid line graphs the approximation $\phi(t) = e^t + c_2 e^{2 t}$.
The long dashed line graphs $\phi(t) = e^t + c_2 e^{2t} + c_3
e^{3t}$. The approximate solutions computed up to $e^{4t}$, $e^{5t}$
and $e^{6 t}$ are very close in this range of $t$ and are all
represented in the short dashed line. One can see that after going
through the first turnaround point with coordinates $(t, \phi
(t))\sim$ (1.27, 1.8) the solution decreases, reaching the second
turnaround at around $(t, \phi (t))\sim$ (3.9, -81). The function
$f( \phi (t)) = \text{sign}(\phi (t))\log(1+|\phi(t)|)$ is graphed
to show both turnaround points clearly on the same scale. }
\label{fig:solplot} } The $L = 2$ approximation to the solution for
the tachyon field is
\begin{multline}
\phi(t) \cong
 e^t - \frac{64\,e^{2\,t}}{243\,{\sqrt{3}}} + 0.002187\,e^{3\,t} - \\
  3.9258\,{10}^{-6}\,e^{4\,t} + 4.9407\,{10}^{-10}\,e^{5\,t} +
  6.3227\,{10}^{-12}\,e^{6\,t}.
\label{eq:solution}
\end{multline}
Plotting the result we observe that for small enough $t$ the term
$e^{t}$ dominates and the solution decays as $e^{t}$ at $-\infty$.
Then, as $t$ grows, the second term in (\ref{eq:solution}) becomes
important.  The solution turns around and $\phi(t)$ becomes negative,
with the major contribution coming from $e^{2t}$.  Then the next mode,
$e^{3t}$ becomes dominating and so on.  The solution $\phi(t)$ around
the first two turnaround points is shown on the figure
\ref{fig:solplot}. Note that the trajectory passes through the minimum
of the static potential, which is at $\phi \sim 0.545$
\cite{Sen:1999nx, Moeller-Taylor}, well before the first turnaround
point.  \FIGURE { \includegraphics{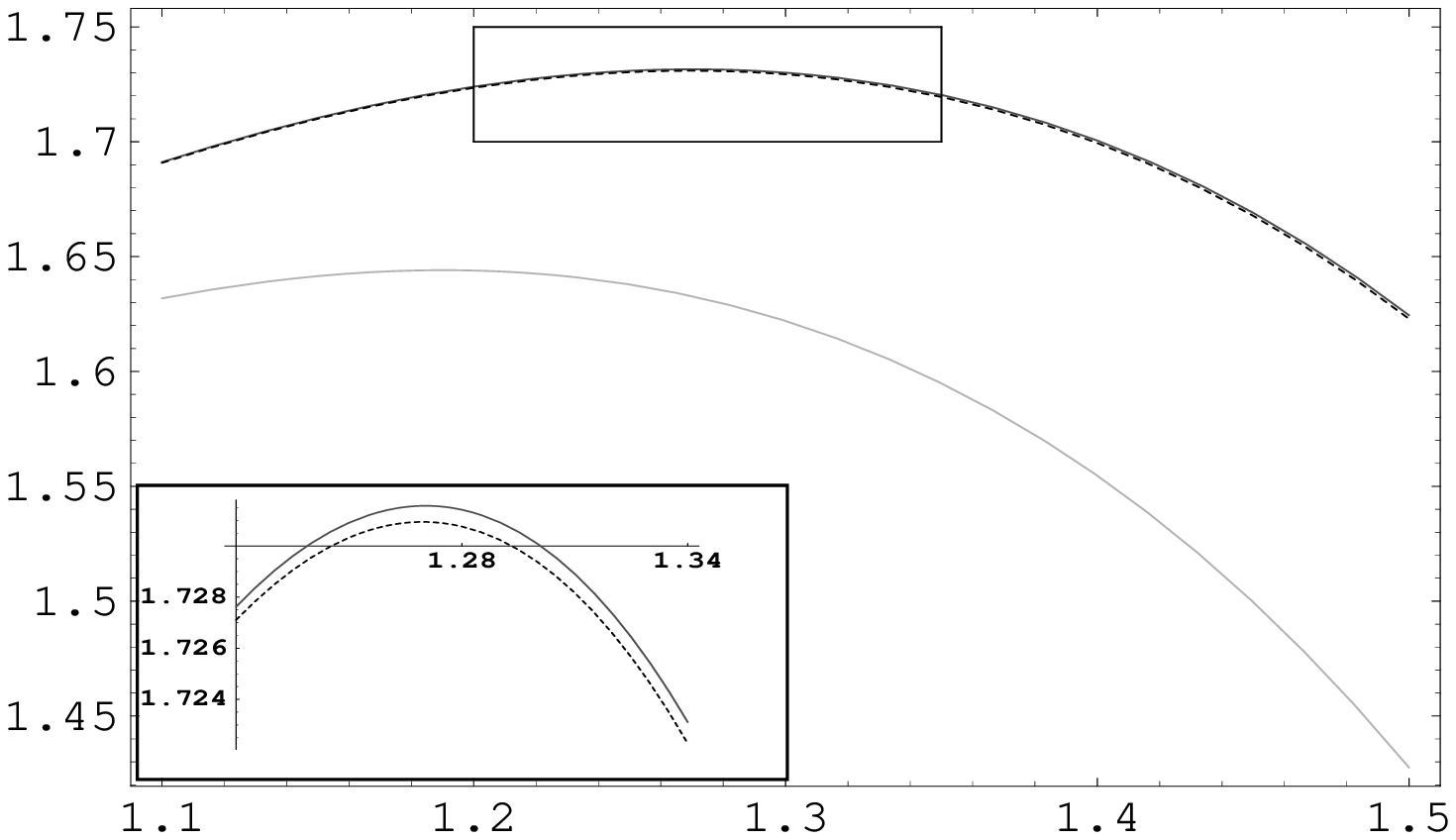}
\caption[x]{\footnotesize First turnaround point for the solution in
$L=2$ truncation scheme.  The large plot shows the approximations
with $\phi^3$ (the gray line), $\phi^4$ (black solid), and $\phi^5$  (dashed
lines) terms in the action taken into account. The smaller plot
zooms in on the  approximations with $\phi^4$  and
$\phi^5$ terms taken into account. The corrections
from higher powers of $\phi$ are very small and the corresponding
plots are indistinguishable from the one of the $\phi^5$
approximation.} \label{fig:firstturn} } \FIGURE {
\includegraphics{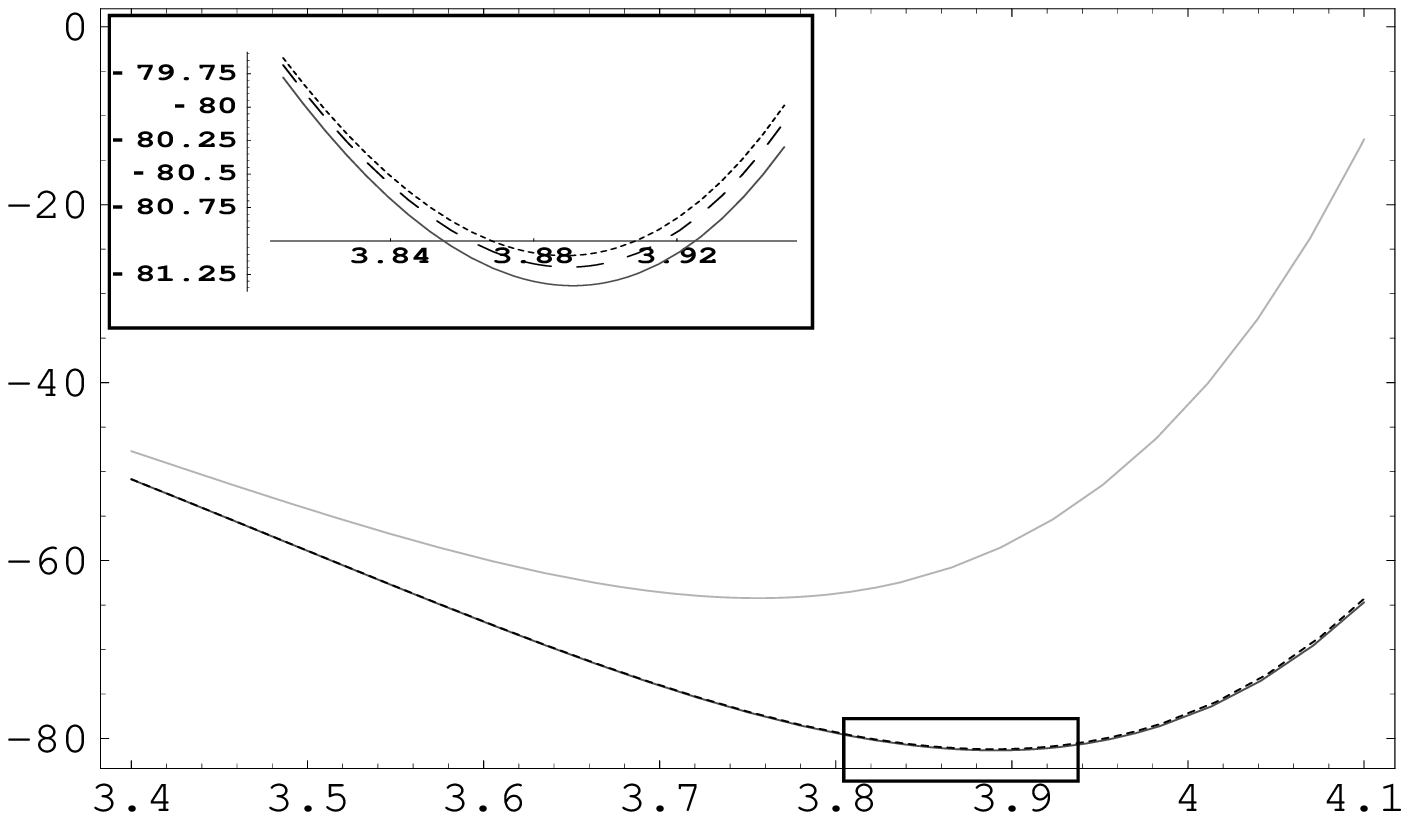}
\caption[x]{\footnotesize Second turnaround point for the solution
in $L=2$ truncation scheme.  The gray line on the large plot shows
the solution computed with the effective action including terms up
to $\phi^4$. The black solid and dashed lines represent higher order
corrections. On the small plot the solid line includes $\phi^5$
corrections, the dotted line includes corrections from the $\phi^6$
term and the dashed line takes into account the $\phi^7$ term.}
\label{fig:secondturn} }

The positions of the first $2$ turnaround points are quite
accurately determined by taking into account the effective action
terms up to $\phi^{5}$. The inclusion of the higher order terms in
the action changes the position of the first $2$ turnaround points
only slightly. Figures \ref{fig:firstturn} and \ref{fig:secondturn}
illustrate the dependence of the position of the first two
turnaround points on the powers of $\phi$ included in effective
action. We interpret these results as strong evidence that, at least
for the effective action at truncation level $L = 2$, the solution
(\ref{eq:expand-solution}) is given by a perturbative series in
$e^{t}$ which converges at least as far as the second turnaround
point, and plausibly for all $t$.

\subsection{Convergence of level truncation}
\label{sec:levelup}

From the results of the previous subsection, we have confidence that
the first two points where the tachyon trajectory turns around are
well determined by the $\phi^4$ and $\phi^5$ terms in the effective
action.  To check whether these oscillations are truly part of a
well-defined trajectory in the full CSFT, we must check to make sure
that the turnaround points are stable as our level of truncation is
increased and the terms in the effective action are computed more
precisely.  From previous experience with level truncated
calculations of the static effective tachyon potential and the
vector field effective action
\cite{Moeller-Taylor,WT-perturbative,cst}, where coefficients in the
effective action generically converge well, with errors of order
$1/L$ at truncation level $L$, we expect that the full tachyon
effective action will also converge well and will lead to convergent
values of $c_n$ within a factor of order 1 of the $L = 2$ results
computed explicitly.

We have computed the $\phi^4$ term in the effective action at levels
of truncation up to $L = 16$.  The results of this computation for the
approximate trajectory $\phi (t)$ are shown in Figure
\ref{fig:levelconv}, which demonstrates the behavior of the first
turnaround point as we include higher level fields.  This computation
shows that the first turnaround point is already determined to within
less than $1\%$ by the level $L = 2$ truncation.  This turnaround
point is also in close agreement with the computations of
\cite{Moeller-Zwiebach}.\footnote{Barton Zwiebach has pointed out that
the position of the first turnaround point for the $\cosh (n t)$
solution of \cite{Moeller-Zwiebach} is very close to the first
turnaround point of the $e^{nt}$ solution which we have computed here,
and that comparing results with two terms in the expansion gives
agreement to within $1\%$.} We take this computation as giving strong
evidence that this turnaround point is real.  We expect from analogy
with other level truncation computations of effective actions and
potentials that the other terms in the effective action considered
here will also generally converge well. Combining the explicit result
for the $\phi^4$ term at high levels of truncation with the
computation of the previous subsection, we have (to us) convincing
evidence that the perturbative expansion $e^{nt}$ for the rolling
tachyon solution is valid well past the first turnaround point, and
that the level truncation procedure converges to a trajectory
containing this turnaround point. Extrapolating the results of this
computation, we believe that the qualitative phenomenon of wild
oscillations revealed by the level $L = 2$ computation is a correct
feature of the time-dependent tachyon trajectory in CSFT, and that
more precise calculations at higher level will only shift the
positions of the turnaround points mildly, leaving the qualitative
behavior intact. \FIGURE { \label{fig:levelconv}
\includegraphics{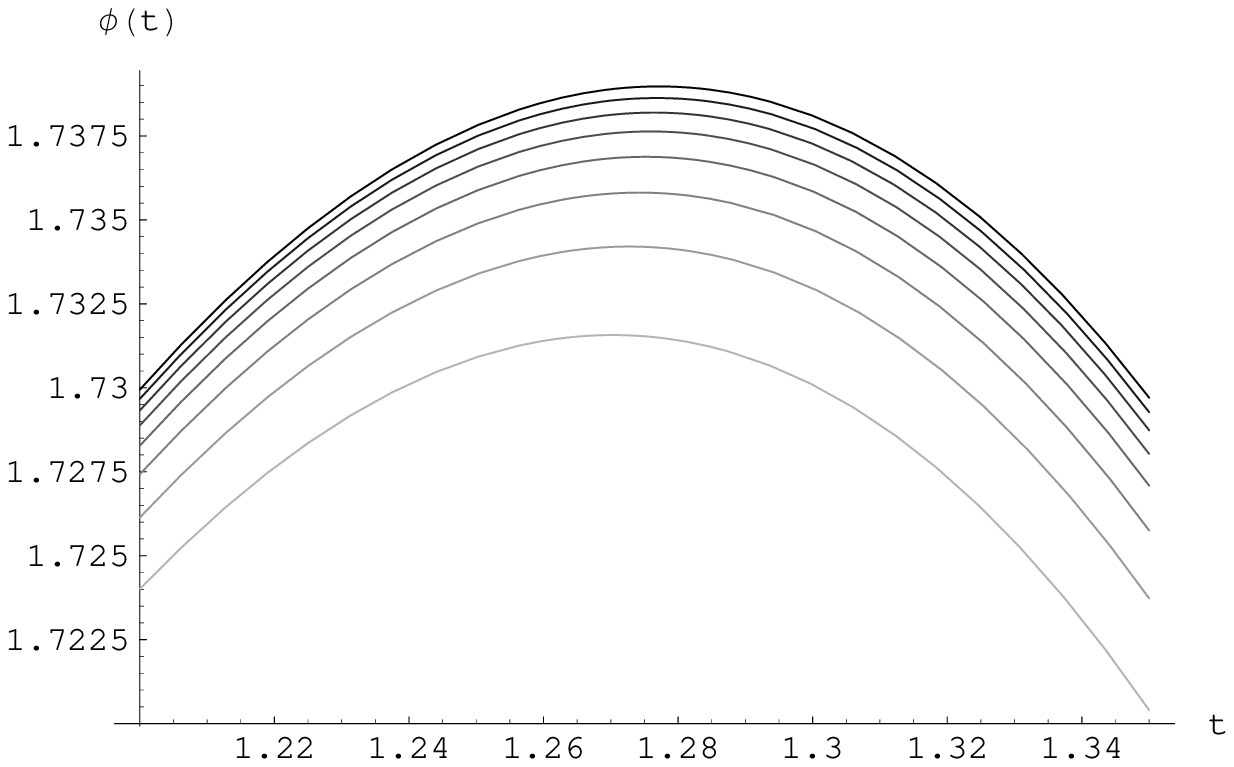} \caption[x]{\footnotesize The figure
shows the convergence of the solution around the first turnaround
point as we increase the truncation level. Bottom to top the graphs
represent the approximate solutions computed with the effective action
computed up to $\phi^4$ and truncation level increasing in steps of 2
from $L = 2$ to $L = 16$. We observe that the turnaround point is
determined to a very high precision already at the level 2.  Similar
behavior is observed for the second turnaround point. } } It is
interesting to compare the behavior of the perturbative expansion of
this time-dependent tachyon solution with a perturbative expansion of
the effective tachyon potential $V (\phi)$.  As found in
\cite{Moeller-Taylor}, the power series expansion for $V (\phi)$ fails
to converge beyond $|\phi | \sim 0.1$ due to a branch point
singularity at negative $\phi$ where the Feynman-Siegel gauge choice
breaks down \cite{Ellwood-Taylor-B}.  Although the potential can be
continued for positive $\phi$ past the radius of convergence using the
method of Pad\'e approximants \cite{Taylor-Pade}, another branch point
associated with the breakdown of Feynman-Siegel gauge is encountered
at a positive $\phi$ just past the minimum near $\phi \sim 0.545$.
Because of these branch points, the expansion for the effective
potential is badly behaved past these points; unlike the
time-dependent solution we have studied here, there is no sense in
which the potential $V (\phi)$ converges for a general fixed value of
$\phi$.  While we initially thought that the wild oscillations of the
low-level computation of the tachyon trajectory $\phi (t)$ might
indicate a similar breakdown of the perturbative expansion, our
results at higher levels seem to give conclusive evidence that this is
not the case.  This suggests that the Feynman-Siegel gauge choice is
valid in the region of field space containing the trajectory $\phi
(t)$ for all $t$ even though the corresponding static $\phi$ lies
outside the region of gauge validity.

\section{Taming the tachyon with a field redefinition}
\label{sec:redef}

Now that we have confirmed that CSFT gives a well-defined but highly
oscillatory time-dependent solution, we want to understand the physics
of this solution.  Although the oscillations seem quite unnatural from
the point of view of familiar theories with only quadratic kinetic
terms and a potential, the story is much more subtle in CSFT due to
the higher-derivative terms in the action. For example, while the
tachyon field apparently\footnote{This is suggested by the effective
tachyon potential at low levels of truncation, which is well-defined
into the region where $V (\phi) > 0$; due to a breakdown of
Feynman-Siegel gauge at large {\em constant} positive $\phi$ at higher
levels of truncation, as mentioned at the end of Section 3, the static
potential is not well-defined in the region of $\phi$ encountered by
the rolling solution in this gauge.} rolls into a region with $V
(\phi)\gg V (0) =0$, the energy of the perturbative rolling tachyon
solution we have found is conserved, as we have verified by a
perturbative calculation of the energy including arbitrary derivative
terms, along the lines of similar calculations in
\cite{Moeller-Zwiebach}.

To understand the apparently odd behavior of the rolling tachyon in
CSFT, it is useful to consider a related story.  In \cite{cst} we
computed the effective action for the massless vector field on a
D-brane in CSFT by integrating out the massive fields.  The
resulting action did not take the expected form of a Born-Infeld
action, but included various extra terms with higher derivatives
which appeared because the degrees of freedom natural to CSFT are
not the natural degrees of freedom expected for the CFT on a
D-brane, but are related to those degrees of freedom by a
complicated field redefinition with arbitrary derivative terms.  In
principle, we expect such a field redefinition to be necessary any
time one wishes to compare string field theory computations (or any
other background-independent formulation) with CFT computations in
any particular background.  The necessity for considering such field
redefinitions was previously discussed in \cite{Ghoshal-Sen,David}.

Thus, to compare the complicated time-dependent trajectory we have
found for CSFT with the marginal $e^{t}$ perturbation of the
boundary CFT found in \cite{Sen:2002nu, Sen:2002in}, we must relate
the degrees of freedom of BSFT and CSFT through a field redefinition
which can include arbitrary derivative terms.  Given an explicit
form $S [T] $ for the BSFT effective tachyon action which admits a
solution $T (t) = e^{t}$, we can construct a perturbative field
redefinition $\phi (t)= \Phi (T (t))$ which maps the BSFT effective
action $S [T]$ to the CSFT effective tachyon action $S [\phi]$.
Since such a field redefinition must map a solution of the field
equations in one picture to a solution in the other picture, it
follows that this map takes the BSFT solution $T (t) =e^{t}$ to the
perturbative solution $\phi (t)$ of the CSFT effective action. In
this section we use an explicit formulation of the BSFT effective
action to compute the leading terms of the field redefinition
relating the effective field theories for $T(\phi)$, the tachyon
field in boundary string field theory and $\phi$, the tachyon in
cubic string field theory. This computation shows in a concrete
example how the complicated dynamics we have found for the tachyon
in CSFT maps to the simple dynamics of BSFT associated with the
marginal deformation $e^{t}$.

In our explicit computations, we use the effective tachyonic action of
BSFT computed up to cubic order in \cite{Coletti:2004ri}; another
approach to computing the BSFT action which may apply more generally
was developed in \cite{Koji-Terashima}.  As we have just discussed, we
expect that a similar field redefinition can be constructed for any
BSFT effective tachyon action.  The BSFT action is determined via the
partition function for the boundary SFT and the tachyon's beta
function.  Thus the particular form of the action depends on the
renormalization scheme for the boundary CFT. The BSFT tachyon $T$ we
use here is, therefore, the renormalized tachyon with the
renormalization scheme of \cite{Coletti:2004ri}.  We now proceed to
construct a perturbative field redefinition relating the CSFT and BSFT
effective actions.  We then will
check explicitly that the field redefinition maps the rolling tachyon
solution $T(t) = e^t$ to the leading terms in the perturbative
solution $\phi(t) = e^t - \frac{64}{243\,{\sqrt{3}}}e^{2\, t}+ \cdots$ which
we have computed in the previous section.  The fact that the field
redefinition is nonsingular at $T = e^t$ is consistent with the Ansatz
$\sum_n c_n e^{nt}$ for the rolling tachyon solution in CSFT.

In parallel with (\ref{eq:effaction}) we write the action for the
boundary tachyon $T$ as
\begin{equation}
S [T] = \sum_n \frac{g^{n-2}}{n!} \int \prod_{i=1}^n ( 2\pi\, d w_i)
\delta\bigl(\sum_i w_i\bigr) \Xi_n^{BSFT}(w_1, \dots, w_n) T (w_1)
\dots T(w_n)
\end{equation}
where the functions $\Xi_n^{BSFT}(w_1, \dots, w_n)$ define the
derivative structure of the term of $n$'th power in $T$.  The kernel
for the quadratic terms is
\begin{equation}
\Xi^{\text{BSFT}}_2(w_1, w_2) = \frac{\G(2 - 2w_1 w_2)}{\G^2(1 - w_1
w_2)}.
\end{equation}
where $\G$ is the Euler gamma function. Denoting $a_1 = - w_2 w_3$,
$a_2 = - w_1 w_3$, $a_3 = - w_2 w_3$ the kernel for the cubic term
can be written as
\begin{align}
\Xi^{\text{BSFT}}_3(w_1, w_2, w_3) =\, & 2(1 + a_1 + a_2 + a_3)
I(w_1, w_2, w_3) + J(w_1, w_2, w_3)
\end{align}
where functions $I(a_1, a_2, a_3)$  and $J(a_1, a_2, a_3)$ are defined by
\begin{align}
I(a_1, a_2, a_3) &= \frac{\G(1 + a_1 + a_2 + a_3)\G(1 + 2 a_1)
\G(1 + 2 a_2)\G(1 + 2 a_3)}{\G(1+ a_1)\G(1 + a_2)\G(1 + a_3)
\G(1 + a_1 + a_2) \G(1 + a_1 + a_3) \G(1 + a_2 + a_3)}, \nonumber \\
J(a_1, a_2, a_3) &=  - \frac{\G(1+ 2 a_1)\G(2+2 a_2 + 2 a_3)}{\G^2(1+a_1)\G^2(1+a_2 +a_3)} +
\text{cyclic}.
\end{align}

We are interested in the field redefinition that relates $S[T]$ with
the CSFT action $S[\phi]$ given in (\ref{eq:effaction}),
(\ref{eq:quadcsftXi}), (\ref{eq:cubcsftXi}). A generic
time-dependent field redefinition can be written in momentum space
as
\begin{multline}
\label{eq:redefinition} \phi(w_1) = \int d w_2\,  \delta(w_1 -
w_2)f_1(w_1, w_2) T (w_2) + \\ \int d w_2\, d w_3\, f_2 (w_1, w_2,
w_3) T(w_2)T(w_3) \delta(w_1 - w_2 - w_3) + \dots\ .
\end{multline}
Note that adding to $f_2$ a term antisymmetric under exchange of
$w_2$ and $w_3$  does not change the field redefinition. Thus, we can
choose $f_2$ to be symmetric under $w_2 \leftrightarrow w_3$.

 The requirement that this field redefinition maps the CSFT action to the
BSFT action,
\begin{equation}
S[\phi(T)] = S[T],
\end{equation}
imposes conditions on the functions $f_i (w_1, \ldots,w_{i +1})$. In
order to match the quadratic terms, $f_1$ must satisfy the equation
\begin{equation}
\label{eq:f1eq} \Xi_2^{\text{BSFT}}(w_1, w_2) - f_1(w_1, w_1)
f_1(w_2, w_2)\Xi_2^{CSFT}(w_1, w_2) \approx 0.
\end{equation}
In this equation the approximate sign means that the left hand side
becomes equal to the right hand side when inserted into $\int dw_1\,
dw_2\, \delta(w_1 + w_2) \phi(w_1) \phi(w_2)$ for arbitrary
$\phi(w)$.\footnote{ When matching the quadratic terms this
condition implies strict equality since both $\Xi_2$'s are
symmetric, but in general the condition is less restrictive.
Considering a discrete analogue, it is easy to see that the equation
$M_{kl} c_k c_l = 0$ is equivalent to $M_{kl} + M_{lk} = 0$.
Similarly,  the equation $M_{n_1, \dots, n_k} c_{n_1} \dots c_{n_k}$
= 0 is equivalent to the sum over permutations $\sigma$ on $n$
elements
 $\sum_{\sigma} M_{\sigma(n_1,\ldots,n_k)} = 0$.} Solving
equation (\ref{eq:f1eq}) we find
\begin{equation}
\label{eq:f1f2} f_1(w, w)\, \equiv f_1(w) = \, \sqrt{\frac{1}{1+w^2}
\frac{\G(2 + 2 w^2)}{\G^2(1+w^2)}}.
\end{equation}
The analogous equation for $f_2$ is
\begin{align}
\label{eq:f2eq}
\nonumber
 \frac{1}{3}\, \Xi_3^{\text{BSFT}}(w_1, w_2, w_3) \approx \, \frac{1}{3}\, &f_1(w_1) f_1(w_2) f_1(w_3) \Xi_3^{CSFT}(w_1, w_2, w_3) + \\
 & f_1(w_1) f_2(-w_1, w_2, w_3) \Xi_2^{CSFT}(- w_1, w_1).
\end{align}
In constructing a consistent
perturbative field redefinition, we further require that
the field redefinition must map the mass-shell states
correctly, by keeping the mass-shell component of any $\phi(w)$
intact.  In other words the mass-shell component of the Fourier
expansion of $\phi(t)$
should not be affected by the higher-order terms $f_2, etc.$
This translates to a
restriction on $f_2$
\begin{equation}
\label{eq:f2constr} f_2(- w_1, w_2, w_3)|_{w_1^2 = -1} = 0.
\end{equation}
This constraint is crucial for the field redefinition to correctly relate
the on-shell scattering amplitudes for $T$ with those for $\phi$.
It also ensures that the solution of the classical equations of motion for $T$
maps to the solution of the equations of motion for $\phi$.

Equation (\ref{eq:f2eq}) can be simplified by making a substitution
\begin{eqnarray}
\lefteqn{ f_2(- w_1, w_2, w_3) =}\label{eq:f2andA} \\[0.1in] &  &
\frac{\Xi_3^{\text{BSFT}}(w_1, w_2, w_3)/f_1(w_1) -
\Xi_3^{CSFT}(w_1, w_2, w_3) f_1(w_2) f_1(w_3) }{ \Xi_2^{CSFT}(-w_1,
w_1)}  A_2(w_1, w_2, w_3) \nonumber
\end{eqnarray}
giving a simple equation for $A_2(w_1, w_2, w_3)$
\begin{equation}
A_2(w_1, w_2, w_3) \approx \frac{1}{3}.
\end{equation}
Thus, we would now like to find a function $A(w_1, w_2, w_3)$ on the momentum
conservation hyperplane $-w_1+w_2+w_3 = 0$, symmetric (by choice)
under the exchange
of $w_2$ and $w_3$ and satisfying
\begin{equation}
\label{eq:acond} A_2(w_1, w_2, w_3) + A_2(w_2, w_3, w_1) + A_2(w_3,
w_1, w_2) = 1,
\end{equation}
with the constraint\footnote{One can check that the prescription used
here is correct on a simple example.  The simplest example is a
polynomial system with a finite number of degrees of freedom and no
time dependence.  For a system with time-dependence, consider mapping
the action of the harmonic oscillator to the action of an anharmonic
oscillator with a cubic potential term $- \frac{1}{3}\phi^3$.  With
the choice of $A$ that preserves the mass-shell modes one gets a field
redefinition that correctly maps the solution of the harmonic
oscillator $e^{it}$ to the perturbative solution of the anharmonic
oscillator $e^{it}-\frac{1}{3}e^{2it}+ \dots$. Attempting to choose,
for example, a completely symmetric $A$
gives rise to an unwanted additional factor of $1/3$.}
\begin{equation}
\label{eq:aconstr}
 A_2(w_1, w_2, w_3)|_{w_1^2 = -1} = 0.
\end{equation}
It is sufficient for our needs here to consider a discrete case,
where $w_1$, $w_2$, $w_3$ are (imaginary) integers. Indeed, as we
are expanding in powers of $e^{t}$, we are restricting attention to
fields expressed in modes with $w = in$. It is easy to check that
the discretized form of $A$ given by
\begin{equation}
\label{eq:A-cont} A(w_1, w_2, w_3) =
\begin{cases}
\frac{1}{3}, & w_{1,2,3} \neq  \pm i \\
 0, & w_1 = \pm i \\
 \half,  & w_2 = \pm i,\ w_{1,3} \neq  \pm i \quad \text{or} \quad  w_3 = \pm i,\ w_{1,2} \neq  \pm i\\
\frac{1}{3}, & w_1 = -2 i,\ w_{2,3} = i \quad \text{or} \quad  w_1 = 2 i,\ w_{2,3} = -i\\
 1 & w_1 = 0,\ w_2 = - w_3 = \pm i
\end{cases}
\end{equation}
is a solution to (\ref{eq:acond}), (\ref{eq:aconstr}). Of course, to
define a consistent field redefinition for the complete field theory
for all functions $\phi$ on $t \in (-\infty, \infty)$ we would need
to construct a continuous function $A$, satisfying the above
conditions. Since this is not crucial for the development of this
paper we relegate a brief discussion of the construction of such a
function to Appendix \ref{app:B}.

Let us make a few comments on the field redefinition.
\begin{itemize}
\item
While $f_1 (w)$ is smooth at the mass-shell point due to a
cancelation of poles at $w^2 = -1$, there is a pole at $w^2 = -3/2$
below which the expression under the square root becomes negative.
This means that the field redefinition (\ref{eq:redefinition}) is
only well defined on the subspace  $T (w)$ with $w^2 > - 3/2$.
Within this region $f_1(w)$ is smooth without any zeroes or poles.
The mass-shell point, $w^2 = -1$ lies within this region.  Related
observations were made in \cite{Koji-Terashima}.
\item
The function $A_2$ represents a universal part of $f_2$ and is
independent of the particular properties of the CSFT and BSFT
actions. For example to map the action of a harmonic oscillator to
the action of an anharmonic oscillator we could use the same $A$.
\item
The term multiplying $A_2(w_1, w_2, w_3)$ in (\ref{eq:f2andA}) has a
number of poles. However it is non-singular in two important cases.
The first case is for spatially dependent fields with $k_{1, 2, 3}^2
= 1$, when the tachyon fields in both frames $T(k)$ and $\phi(k)$
are on mass-shell. At this point the two summands in the numerator
of (\ref{eq:f2andA}) cancel, and there is no pole at this point. The
requirement of this cancelation was used in \cite{Coletti:2004ri} to
fix the normalization of BSFT action.

The second case is the one of the rolling tachyon.  In this case
$T(w_2)$ and $T(w_3)$ are on mass-shell: $w_2^2 = w_3^2 = - 1$,
while $\phi(w_1)$ is not: $w_1^2 = -4$. There is a potential
singularity in the term $\Xi_3^{\text{BSFT}}(w_1, w_2,
w_3)/f_1(w_1)$ in the numerator. $f_1(w_1)$ has a zero at $w_1^2 =
-4,$ but the functions $I$ and $J$ in the  $\Xi_3^{\text{BSFT}}$
have a stronger zero resulting in a zero at that point.
\end{itemize}

Finally, we want to check that the field redefinition maps the
rolling tachyon solution of BSFT into the perturbative solution that
we have found in section \ref{sec:solvingeom}. Plugging the rolling
solution $T_{\text{rolling}}(t) = e^{t}$ into the field redefinition
and computing the numerical values we obtain
\begin{equation}
\phi(t) =  e^t - \frac{64\,e^{2\,t}}{243\,{\sqrt{3}}} + \dots
\end{equation}
which exactly reproduces the leading order terms in the perturbative
CSFT solution found in section \ref{sec:solvingeoms}.  As we include
higher powers of $\phi$ in the field redefinition we should continue
to generate the higher power terms $e^{nt}$ in the perturbative
solution.

\section{Discussion}
\label{sec:discussion} In this paper we have confirmed and expanded
on the earlier results of \cite{Moeller-Zwiebach} and
\cite{Fujita-Hata}, which suggested that in CSFT the rolling tachyon
oscillates wildly rather than converging to the stable vacuum.  We
have shown that the oscillatory trajectory is stable when
higher-level fields are included and thus correctly represents the
dynamics of CSFT.  We have found that the energy of this oscillatory
solution is conserved.  We have further shown that this dynamics is not
in conflict with the more physically intuitive $e^{t}$ dynamics of
BSFT by explicitly demonstrating a field redefinition, including
arbitrary derivative terms, which (perturbatively) maps the CSFT
action to the BSFT action and the oscillatory CSFT solution to the
$e^{t}$ BSFT solution.

This resolves the outstanding puzzle of the apparently different
behavior of the rolling tachyon in these two descriptions of the
theory.  On the one hand, this serves as further validation of the
CSFT framework, which has the added virtue of background-independence,
and which has been shown to include disparate vacua at finite points
in field space.  On the other hand, the results of this paper serve as
further confirmation of the complexity of using the degrees of freedom
of CSFT to describe even simple physics.  Further insight into the
physical properties of the solution we have computed here, such as an
understanding of the pressure of the rolling tachyon field, would
require new insight or substantial computation.  As noted in previous
work, many phenomena which are very easy to describe with the degrees
of freedom natural to CFT, such as marginal deformations
\cite{Sen-Zwiebach}, and the low-energy Yang-Mills/Born-Infeld
dynamics of D-branes \cite{cst} are extremely obscure in the variables
natural to CSFT.  This is in some sense possibly an unavoidable
consequence of attempting to work with a background-independent
theory: the degrees of freedom natural to any particular background
arise in complicated ways from the underlying degrees of freedom of
the background-independent theory. This problem becomes even more
acute in the known formulations of string field theory, which require
a canonical choice of background to expand around, when attempting to
describe the physics of a background far from the original canonical
background choice, such as when describing the physics of the true
vacuum using the CSFT defined around the perturbative vacuum
\cite{Ellwood-Taylor-A, Ellwood-Taylor-B}.  The complexity of the
field redefinitions needed to relate even simple backgrounds such as
the rolling tachyon discussed in this paper to the natural CFT
variables make it clear that powerful new tools are needed to take
string field theory from its current form to a framework in which
relevant physics in a variety of backgrounds can be clearly computed
and interpreted.

\section*{Acknowledgements}
We are grateful to Dmitry Belov, Ian Ellwood, Ted Erler, Guido
Festuccia, Gianluca Grignani, David Gross, Hiroyuki Hata, Koji
Hashimoto, Satoshi Iso, Yuji Okawa, Antonello Scardicchio, Ashoke Sen,
Jessie Shelton, and Barton Zwiebach for discussion and useful
comments.  This work was supported by the DOE under contract
\#DE-FC02-94ER40818.

\vspace*{0.2in}

\appendix

\section{Perturbative computation of effective action}
\label{app:A} \vspace*{0.1in}

We have used two methods to compute the coefficients in the effective
action $S [\phi (t)]$.  The first method, as described in the main
text, consists of solving the equations of motion for each field
perturbatively in $\phi$.  The second method consists of computing the
effective action by summing diagrams which can be computed using the
method of level truncation on oscillators.  This approach is
summarized briefly
here, and applied to the computation of the term of order $\phi^4$ in
the effective action.


The classical effective action for the tachyon can be perturbatively
computed as a sum over all tree-level connected Feynman diagrams.
\FIGURE{\includegraphics{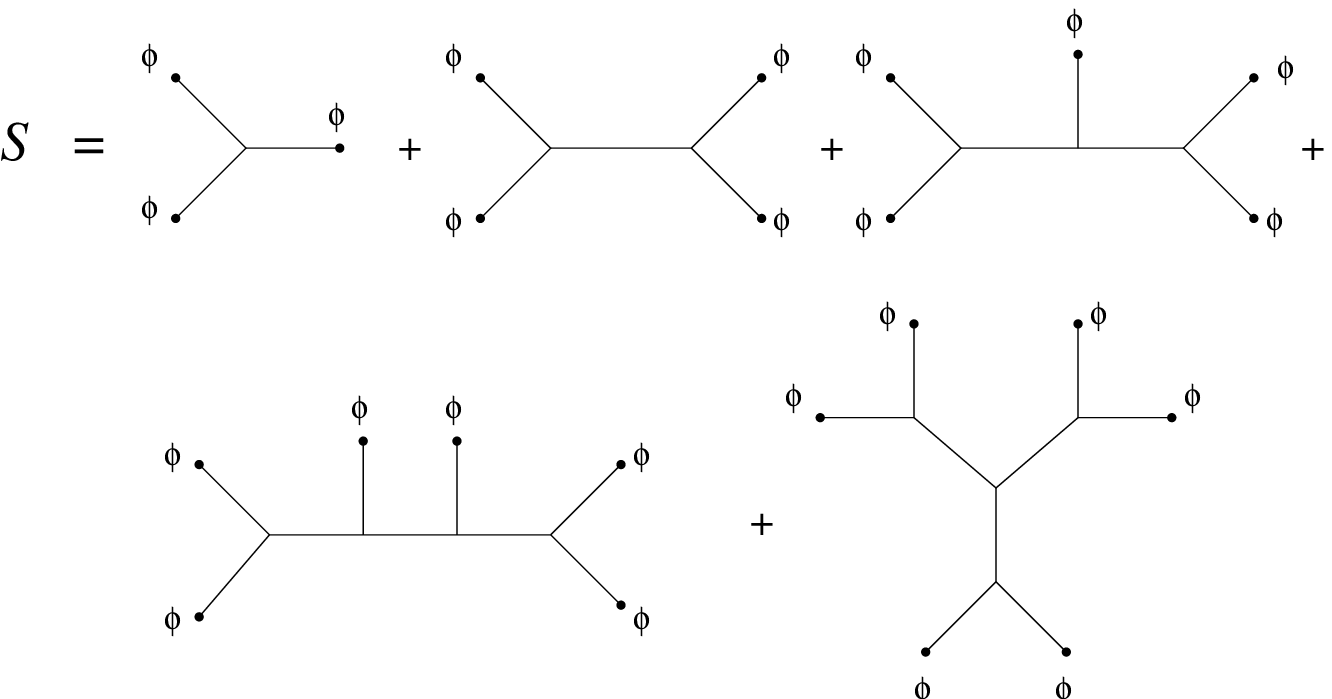} \caption[x]{\footnotesize The
first few diagrams contributing to the effective action}
\label{fig:diagrams}}

A method for computing such diagrams to high levels of truncation in
string field theory was presented in \cite{WT-perturbative}, and
used in \cite{cst} to compute the effective action for the massless
vector field.  A review of this approach is given in
\cite{WT-perturbative-review}.  Using this method, the contribution
of a given Feynman diagram with $n$ vertices, $n-1$ propagators and
$n+2$ external fields is given by an integral of the form
\begin{equation}
\label{eq:actioncontrib} \delta S = \int \prod_{i = 1}^{n} {d k_i}
(2 \pi)^d \delta(\sum k_i) \int \prod_{j=1}^{n-3} \frac{d\,
\sigma_j}{\sigma_j^2} \Det \left( \frac{1 - \cX \cP}{ (1 - \cV
\cP)^{13}} \right) \exp\left(k_i \cQ^{ij} k_j \right)\phi(k_1)\dots
\phi(k_j).
\end{equation}
In this formula $\cV$ and $\cX$ are $n \times n$ block matrices whose blocks
are matter and ghost Neumann coefficients $V^{r s}$ and $X^{r s}$  of the cubic string field theory vertex.
More precisely
\begin{align}
\cV &=
\begin{pmatrix}
V^{r_1 s_1} & 0  &\dots & 0   \\
0 & V^{r_2 s_2}  &\dots & 0   \\
\dots & \dots & \dots & \dots \\
0 & 0 &  \dots & V^{r_n s_n}
\end{pmatrix}, &
\cX &=
\begin{pmatrix}
X^{r_1 s_1} & 0  &\dots & 0   \\
0 & X^{r_2 s_2}  &\dots & 0   \\
\dots & \dots & \dots & \dots \\
0 & 0 &  \dots & X^{r_n s_n}
\end{pmatrix}.
\end{align}
When using level truncation $V^{rs}$ and $X^{rs}$ become
$3 L \times 3 L$ matrices of real numbers.
The matrix $\cP$ encodes information about propagators, external states and the graph structure of the
diagram.  We define it as
\begin{equation}
\cP = K^T \hat \cP K.
\end{equation}
Here  $\hat \cP$ is a block-diagonal matrix of the form
\begin{equation}
\hat \cP =
\begin{pmatrix}
P(\sigma_1) & 0  &\dots & 0      & 0 & \dots & 0 \\
0 & P(\sigma_2)  &\dots & 0      & 0 & \dots & 0 \\
\dots & \dots & \dots & \dots    & 0 & \dots & 0 \\
0 & 0 &  \dots & P(\sigma_{n-1}) & 0 & \dots & 0 \\
0 & 0 & \dots & 0 & 0 & \dots & 0  \\
\dots & \dots & \dots & \dots & \dots & \dots &\dots &\\
0 & 0 & \dots & 0 & 0 & \dots & 0
\end{pmatrix}.
\end{equation}
The diagonal blocks $P(\sigma_i)$ correspond to propagators.  In the
level truncation scheme the block $P(\sigma)$ of $\hat \cP$ is the
$2 L \times 2 L$ matrix
\begin{equation}
P(\sigma) =
\begin{pmatrix}
0 & P_{1 2}(\sigma) \\
P_{21}(\sigma) & 0
\end{pmatrix}
\end{equation}
where
\begin{equation}
P_{12}(\sigma) = P_{21}(\sigma) =
\begin{pmatrix}
\sigma & 0 & \dots & 0 \\
 0 & \sigma^2 & \dots & 0 \\
 \dots & \dots & \dots & \dots\\
 0 & 0 & \dots & \sigma^L \\
\end{pmatrix}.
\end{equation}
The last $n + 2$ rows and columns of $\hat \cP$ are filled with
zeroes which correspond to external tachyon states.
\FIGURE{\includegraphics{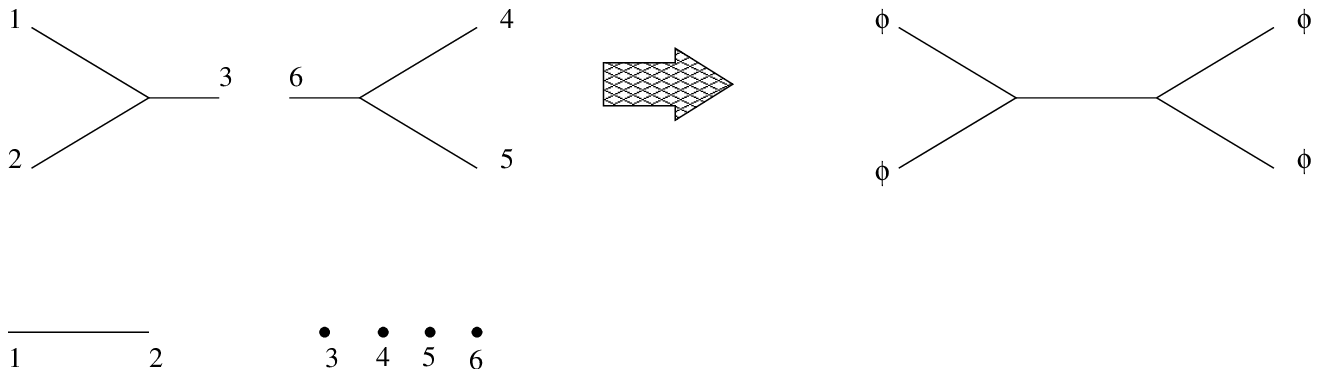} \caption[x]{\footnotesize To
construct the 4 point diagram we label consecutively the edges of
vertices on one hand and propagators and external states on the
other. The matrix $K$ corresponds to a permutation which glues them
into one diagram.} \label{fig:4-point}}

The matrix $K$ is the block permutation matrix that encodes
information on the graph structure of the diagram. The corresponding
permutation $\kappa$ connects the external states and propagators to
vertices as illustrated for the 4-point diagram in Figure
\ref{fig:4-point}. The vertices' edges which are labeled $1$ through
$6$ are connected by permutation to the propagator edges labeled $1$
and $2$ and the external points labeled $3$, $4$, $5$ and $6$. As we
can see a suitable choice of a permutation is
\begin{equation}
\kappa :
\begin{pmatrix} 1 & 2 & 3 & 4 & 5 & 6
\end{pmatrix} \rightarrow
\begin{pmatrix}
3 & 6 & 1 & 2 & 4 & 5
\end{pmatrix},
\end{equation}
which corresponds to
\begin{equation}
K =
\begin{pmatrix}
0\ & 0\ & 1\ & 0\ & 0\ & 0\ \\
0\ & 0\ & 0\ & 0\ & 0\ & 1\ \\
1\ & 0\ & 0\ & 0\ & 0\ & 0\ \\
0\ & 1\ & 0\ & 0\ & 0\ & 0\ \\
0\ & 0\ & 0\ & 1\ & 0\ & 0\ \\
0\ & 0\ & 0\ & 0\ & 1\ & 0\ \\
\end{pmatrix}.
\end{equation}
For example, multiplying matrices for the 4-point amplitude we find
\begin{align}
\cV \cP &= ({\td V}^{11})^2, & \cX \cP &= ({\td X}^{11})^2,
\end{align}
where
\begin{align}
 {\td V}^{11}_{mn} &= \sigma^{\frac{m+n}{2}} V^{11}_{mn},     &  \td X^{11}_{mn} &=
 \sigma^{\frac{m+n}{2}}
 X^{11}_{mn}.
\end{align}

The contribution from the Feynman diagram with 4 external tachyons
is given by \cite{cst}
\begin{multline}
\label{eq:4ampl}
\frac{e^{-3 V_{00}^{11}}g^{2}}{2}
\int \prod_{i = 1}^{4} (2\pi d w_i)\phi(w_i)  \delta(\sum w_i) \\
\int \frac{d\,
\sigma}{\sigma^2} \Det
\left( \frac{1 - (\td X^{11})^{2} }{ [1 - (\td V^{11})^{2} ]^{13}} \right)
\sigma^{-\half[ (w_{1}+w_{2})^{2}+(w_{3}+w_{4})^{2}]}
\exp\left(-w_i \cQ^{ij} w_j\right),
\end{multline}
with $\cQ^{ij}$ defined as
\begin{align}
\cQ^{ij}&=\cU^{i3}_{0.}\frac{1}{1-(\td V^{11})^{2}}\td
V^{11}\cU^{3j}_{.0}+\cU^{ij}_{00}
&   i,j=1,2 &\text{ or } i,j=3,4,\\
\cQ^{ij}&=-\cU^{i3}_{0.}\frac{1}{1-(\td V^{11})^{2}}C\cU^{3j}_{.0} &
(i=1,2 \text{ and } j=3,4) &\text{ or } (i=3,4 \text{ and } j=1,2)
\end{align}
where $\cU^{ij}$ is given by
\begin{equation}
\cU^{ij}=
\begin{pmatrix}
V_{00}^{ij}-V_{00}^{i3}-V_{00}^{3j}+V_{00}^{33} & \ \
\td V_{0n}^{ij}-\td V_{0n}^{3j}\\
\td V_{m0}^{ij}-\td V_{m0}^{i3} & \td V_{mn}^{ij}
\end{pmatrix},
\end{equation}
and $C_{mn}=\delta_{mn}(-1)^{n}$. Considering only the contribution
coming from level 2 fields, we have to consider only these Neumann
coefficients whose powers and products sum up to a total oscillator
level of 2, i.e. $V_{01}, V_{11}, V_{02}$ and $X_{11}$
\cite{WT-perturbative}\footnote{If we want to calculate the quartic
term in the effective action we have to subtract the contribution
from the tachyon in the propagator.}. Doing so equation
(\ref{eq:4ampl}) simplifies a lot and  the integral over the modular
parameter reduces to
\begin{equation}
\int d \sigma \sigma^{-\half\big[(w_{1}+w_{2})^{2}+(w_{3}+w_{4})^{2}\big]}.
\end{equation}
Performing this integral it is easy to get the same result as in
formula (\ref{eq:L2quartic}).


\section{Construction of $A(w_1, w_2, w_3)$ in the continuous case}
\label{app:B}
\FIGURE{\includegraphics{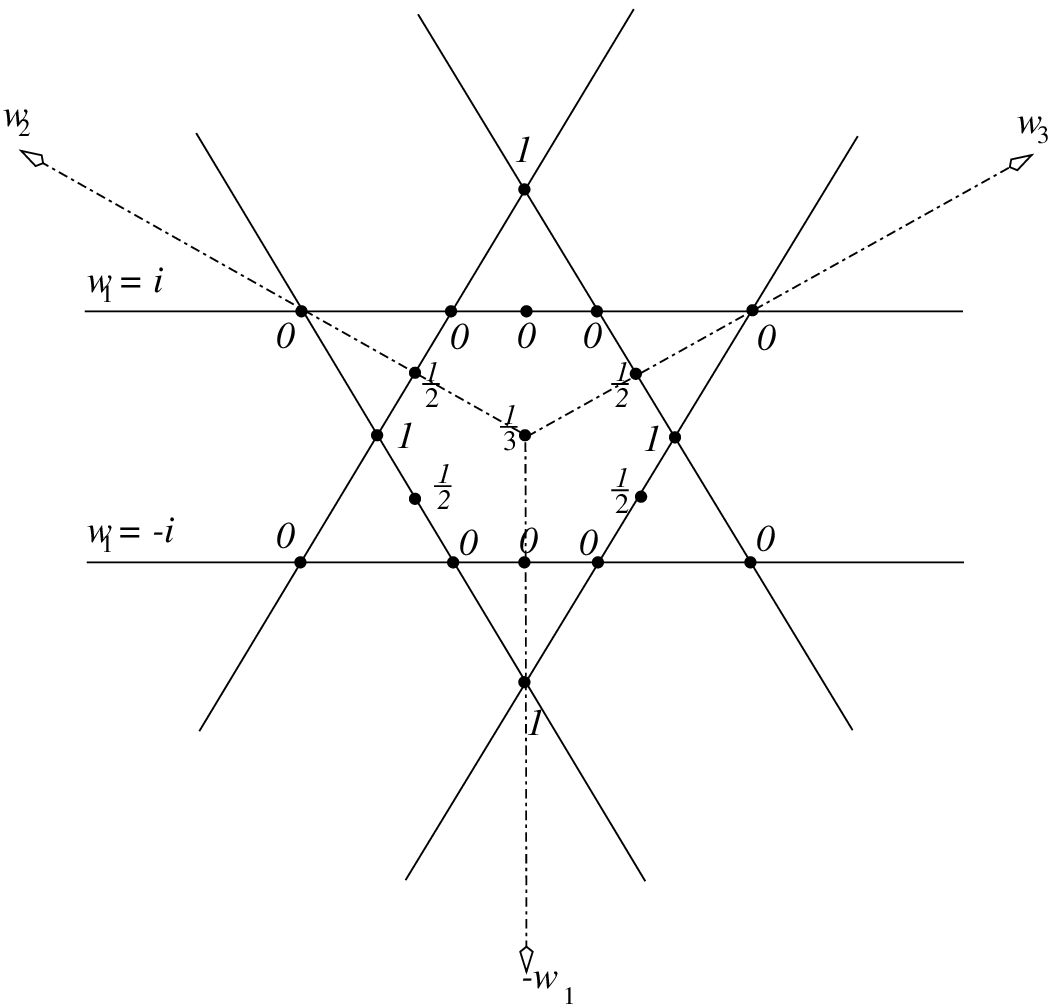}
\caption[x]{\footnotesize Construction of a continuous $A(w_1,
w_2, w_3)$:
The figure shows the plane $-w_1+w_2+w_3 = 0$ coincident with the
plane of the paper. Dashed lines are the coordinate axes $-w_1$,
$w_2$, $ w_3$, going at equal angles out of the plane of the paper.
The two solid horizontal lines are intersections of the plane
$-w_1+w_2+w_3 = 0$ with the planes $w_1 = \pm i$.  According to
(\ref{eq:B2}) the function $A$ is zero along these lines. Clearly,
in this projection the cyclic shift of momenta $w_i$ corresponds to
a 60 degree rotation. Thus, the condition (\ref{eq:B1}) implies that
the sum of the values of $A$ over the vertices of any equilateral
triangle centered at the origin is one.  Together with the
reflection symmetry $w_2 \leftrightarrow w_3$ this allows us to fix the
value of $A$ at several discrete points. The values are shown on the
figure. The slanted solid lines show the locus of the vertices
of equilateral triangles with one vertex fixed on the lines $w_1^2 =
\pm i$.  The assignment of a value for $A$ on one of the slanted lines
defines the values on the other line, related by $60^\circ$
rotation.  These assignments can be made
continuously along the lines while taking the values $0$, $1$, and
$\half$ at the symmetrically positioned points. One can then continuously
extend $A$ into the rest of the plane, while maintaining
(\ref{eq:B1}) by interpolating between the values of $A$
at the boundaries. } \label{fig:Ademo}}
As we have discussed in section \ref{sec:redef}, in order to
construct the field redefinition from BSFT to CSFT that preserves
general solutions to the equations of motion we need a continuous
function $A(w_1, w_2, w_3)$, defined on the plane $-w_1+w_2+w_3 = 0$
and satisfying
\begin{equation}
\label{eq:B1} A_2(w_1, w_2, w_3) + A_2(w_2, w_3, w_1) + A_2(w_3,
w_1, w_2) = 1,
\end{equation} and
\begin{equation} \label{eq:B2}
 A_2(w_1, w_2, w_3)|_{w_1^2 = -1} = 0.
\end{equation}
Figure \ref{fig:Ademo} illustrates the construction of the desired
function.

\end{document}